\documentclass[final,3p,times]{elsarticle}
\usepackage{amsmath,amssymb,bm,graphicx,multicol,color,soul,dsfont,multirow}
\usepackage[colorlinks=true, pdfstartview=FitV, linkcolor=blue, citecolor= blue, urlcolor=blue]{hyperref}
\biboptions{sort&compress}
\allowdisplaybreaks
\journal{Nuclear Physics A}

\renewcommand{\arraystretch}{1.5}

\begin{document}
\begin{frontmatter}
\title{Radiative production and decays of the exotic $\eta_1^\prime(1855)$ and its siblings}

\author[ujk]{Vanamali Shastry\corref{1}}
\ead{vanamalishastry@gmail.com}

\author[ujk,guf]{Francesco Giacosa}
\ead{francescogiacosa@gmail.com}

\cortext[1]{Corresponding author}
\address[ujk]{Institute of Physics, Jan Kochanowski University, ul.  Uniwersytecka  7,  P-25-406  Kielce,  Poland}
\address[guf]{Institute for Theoretical Physics, Johann Wolfgang Goethe - University, Max von Laue--Str. 1 D-60438 Frankfurt, Germany}

\begin{abstract}
We study the production of the light $J^{PC}=1^{-+}$ hybrid isoscalars $\eta_1^\prime(1855)$ and the yet undiscovered $\eta_1^{hyb}(1660)$ as intermediate states in the radiative decays of the charmonium ($J/\psi$) to two conventional mesons using a flavor symmetric Lagrangian.
For this purpose, we use the $J/\psi\to\gamma\eta_1^\prime(1855)\to\gamma\eta\eta^\prime$ process as the reference. We find that some of the decay channels have branching ratios similar to or larger than that of the $\gamma\eta\eta^\prime$ channel and are sensitive to the mixing between the hybrid isoscalars. We propose that relatively stable $\gamma\eta f_1(1285)$ channel be explored for the presence of the light hybrid isoscalar $\eta_1^{hyb}(1660)$. 
We also exploit the strong decay channels containing at least one vector meson to study the radiative decays of the whole hybrid nonet $\{\pi_1(1600),K_1^{hyb}(1750),\eta_1^{hyb}(1660),\eta_1'(1855)\}$. 
We find that the hybrids cannot radiatively decay into the $I=0$ pseudoscalars. Furthermore, the vector decay channels ($(\rho/\omega/\phi)\gamma$) of the hybrid isoscalars are sensitive to the strangeness content of the hybrids. We also provide estimates for the branching fractions for the radiative production and partial widths for the radiative decays of the hybrids.
\end{abstract}

\end{frontmatter}

\section{Introduction}\label{sec:intro}
The study of the radiative decays of the heavy quarkonia into light pseudoscalar mesons has resulted in the observation of a significant number of conventional as well possibly exotic resonances \cite{Workman:2022ynf}. 
The process,
\begin{equation}
    [\bar{Q}Q]_h\to \gamma \mathcal{R}^* \to \gamma\phi_1\phi_2\ldots
\end{equation}
where, $[\bar{Q}Q]_h$ is a heavy quarkonium, $\mathcal{R}^*$ is a resonance constituting the light quarks, and $\phi_i$ are pseudoscalar mesons has been studied extensively across various colliders and accelerators \cite{Workman:2022ynf}. Particularly, the study of the radiative decays of the charmonium $J/\psi$ has resulted in the discovery of numerous scalar and tensor resonances \cite{Ablikim:2006db,BESIII:2013qqz,Dobbs:2015dwa,BESIII:2016fbr,BESIII:2016gkg,BESIII:2018ubj,Sarantsev:2021ein}, including the supernumerary (light) $X$ states \cite{Workman:2022ynf}. Recently, one such process was studied by the BESIII collaboration which resulted in the observation of the $\eta_1'(1855)$ state with quantum numbers $1^{-+}$ \cite{BESIII:2022riz,Ablikim:2022ugk}. This state was observed in the decay process $J/\psi\to\gamma\eta_1'(1855)\to\gamma\eta\eta^\prime$, and the branching ratio for the same was extracted as $(2.70\pm0.41^{+0.16}_{-0.35})\times 10^{-6}$ \cite{BESIII:2022riz,BESIII:2022iwi}. This discovery has spurred the community to reevaluate the light exotic nonets. Various models were proposed to explain the nature of the $\eta_1'(1855)$, ranging from a hybrid state to meson molecule \cite{Chen:2022isv,Shastry:2022mhk,Tang:2021zti,Dong:2022cuw,Qiu:2022ktc,Chen:2022qpd,Wang:2022pin,Yang:2022rck,Wan:2022xkx,Guo:2022xqu,Su:2022eun,Ji:2022blw,Yu:2022wtu,Huang:2022tpq,Su:2022fqr,Shen:2022etd,Wang:2022sib}.\par

Recently, in Ref. \cite{Shastry:2022mhk} we reported, on the basis of a flavor symmetry, parity and charge conjugation (FCP) conserving Lagrangian, the existence of an entire nonet of $1^{-+}$ hybrids in the mass range $1.6-1.9$ GeV. We also found that flavor symmetry arguments lead to a large number of decay channels for these states, of which the ones involving the pseudovector ($1^{+-}$) or axial-vector ($1^{++}$) states are dominant. We use the coupling strengths estimated in this study and a Lagrangian based on flavor symmetry to estimate the width of the $J/\psi\to\gamma\mathcal{R}^*$ decays, and the branching ratios for the $J/\psi\to\gamma\mathcal{R}^*\to\gamma\phi_1\phi_2$, where $\mathcal{R}^*$ is any $1^{-+}$ isoscalar, and $\phi_{1,2}$ are allowed conventional light mesons. This information could prove to be useful for experimental searches for the yet to be discovered light isoscalar belonging to the nonet.\par

In our approach, the hybrid isovector $\pi_1(1600)$ decays dominantly to the $b_1(1235)\pi$ states, in addition to the $\rho\pi$, $K^*K$, $f_1^{(\prime)}\pi$, $\eta^{(\prime)}\pi$, and $\rho\omega$ states \cite{Shastry:2022mhk}, in agreement with the lattice calculations \cite{Woss:2020ayi}. Extrapolating from here, we found that the hybrid kaon (denoted as $K_1^{hyb}(1750)$), with an expected mass of $1.7-1.8$ GeV \cite{Eshraim:2020ucw} and total width of $170-300$ MeV, decays dominantly to the axial kaons $K_1(1270/1400)\pi$, in addition to $a_1(1260)K$ and $b_1(1235)K$. Other modes of the decays of the kaons include the $(\rho/\omega/\phi) K$, $K^*(\pi/\eta)$, $h_1(1170)K$, $\eta^{(\prime)}K$, and $(\rho/\omega)K^*$. 
The two isoscalars of the nonet are expected to be mostly nonstrange and strange due to the homo-chiral nature of the nonet \cite{Shastry:2022mhk,Giacosa:2017pos}. Since only the heavy isoscalar ($\eta_1^\prime(1855)$) has been found so far \cite{BESIII:2022riz}, we estimated the mass of the light isoscalar based on flavor symmetry arguments and homo-chirality and found it to be similar to the mass of the $\pi_1(1600)$, thus we named this putative state as $\eta_1^{hyb}(1660)$. This isoscalar decays dominantly to $a_1(1260)\pi$, in addition to $K^*K$, $\eta\eta^\prime$, and $\rho\rho$ states. Curiously, the light isoscalar is expected to be the narrowest of the hybrids. The heavy isoscalar $\eta_1^\prime(1855)$, on the other hand, has a total width in agreement with the experimental value within the uncertainty and decays dominantly to $K_1(1270)K$ and to $K^*K$, $\eta\eta^\prime$, $f_1\eta$, $K^*K^*$, and $\omega\phi$ states with the width of the last channel being at least an order of magnitude smaller than the rest due to the closeness to the threshold.
Notice that the axial-vector and pseudovector states enter into a large number of the allowed decay channels of the hybrids.\par

The current study is based on the observation that the hybrids decay into vector states quite often (though not dominantly) and hence, can decay radiatively, as per the vector meson dominance (VMD) picture. Transforming the FCP Lagrangian via the VMD transformations, we write down the intended Lagrangian for the radiative decays of the hybrids. The radiative production, on the other hand, is based on the argument that the radiative decays of the $J/\psi$ to hybrids involves the $J/\psi$ meson, a photon, and two gluons converting into a isoscalar state (the isoscalar hybrids in our case), see, Sec. \ref{sec:mod} for details. 
Based on these two Lagrangians, we find that the branching ratios for the radiative production of the hybrid isoscalars in various (final state) channels follows roughly the same trend as the partial widths of their decays into these channels. Further, the $\gamma f_1\eta$ and the $\gamma a_1(1260)\pi$ channels are found to be very sensitive to the angle of mixing between the hybrid isoscalars, see, Sec. \ref{sec:RnD} for details).\par

Quite interestingly, the $J/\psi\to\gamma\eta_1'(1855)\to\gamma\eta\eta^\prime$ decay process was also recently studied on the lattice \cite{Chen:2022isv}. In this study, the authors model the electric and magnetic form factors of the $J/\psi\to\gamma\eta_1'(1855)$ decay process on the lattice, assuming the $\eta_1'(1855)$ to be the non-strange isoscalar hybrid with $J^{PC}=1^{-+}$. Their analysis leads to a width of $8.1\pm 3.3$ MeV for the $\eta_1'(1855)\to\eta\eta^\prime$ decay. We mention here that our model cannot account for such a large mass for the non-strange isoscalar. Hence, we assume that pertinent state is largely made of strange (anti-)quarks and a gluon (see Sec. \ref{sec:mod} for further discussion).\par

In order to set the framework for our study, it is important highlight the current status of the light hybrid mesons as presented in various recent works on the subject, in which different results and interpretations have been discussed. 

The nonrelativistic quark model calculations based on the flux tube model have predicted the existence of one light hybrid nonet with $J^{PC}=1^{-+}$ below $2$ GeV \cite{Isgur:1985vy,Burns:2006wz,Close:1994hc}. These studies, however, are qualitative as they assume a mass of $1.9-2.0$ GeV for the isovectors and the isoscalars which resulted in the total widths of $\sim 200 -350$ MeV. The flux-tube model calculations were subsequently revised by assuming a mass of $1.6$ GeV for the isovector leading to a much smaller width of $\sim 100$ MeV \cite{Page:1998gz}. Interestingly, the possibilities of observing these resonances in the radiative decays of the heavy quarkonia were suggested in Ref. \cite{Isgur:1985vy}. Another study based on the MIT bag model used the masses of the conventional mesons and the then glueball candidate $\iota(1440)$ to extract the (anti)quark and gluon self-energies\cite{Chanowitz:1982qj}. The non-strange hybrids were predicted to have masses around $1.1$ GeV, the kaons and the heavy isoscalar hybrids to be $150$ MeV and $300$ MeV heavier than the isovectors.\par

Field theoretic studies of the mass and decays of the hybrids have been performed using the QCD sum rules, and Dyson-Schwinger/Bethe-Salpeter equations. The analysis of the three-point correlation function in the chiral limit was done using the QCD sum rules in Ref. \cite{Chen:2010ic}. In this study, the mass of the $1^{-+}$ isovector was varied between $1.6$ GeV and $2.0$ GeV. The total width of this state was found to vary significantly with the mass ranging from $200$ MeV (for mass $=1.6$ GeV) to a rather unphysical value of $\sim1250$ MeV (for mass $=2.0$ GeV) \cite{Chen:2010ic}. The non-strange isoscalar, on the other hand was found to be narrower at $60-850$ MeV for the same range of masses. However, the study predicted that the isovector would decay dominantly to the $\rho\pi$ and $f_1(1285)\pi$ states, while the $b_1(1235)\pi$ channel was suppressed by nearly two orders of magnitude. This was in stark contrast to the observations of the flux-tube model, where the hybrids decayed dominantly to the $b_1\pi$ states \cite{Isgur:1985vy}. Further, qualitative predictions were made for the radiative decays of the $J/\psi$ based on a model Lagrangian \cite{Huang:2010dc}.\par

Initial attempts at modeling the hybrid using the Dyson-Schwinger equations with a separable kernel for the quark-antiquark interactions showed the presence of two closely lying hybrid isovectors with $J^{PC}=1^{-+}$ with masses of $1439$ MeV and $1498$ MeV\cite{Burden:2002ps}. But, a subsequent study using the rainbow ladder truncated Bethe-Salpeter equations (RL-BSE) yielded only one hybrid isovector with mass in the range $1.0 - 1.2$ GeV \cite{Krassnigg:2009zh}. Surprisingly, a revised calculation in the same framework resulted in two resonances with masses $1.4$ GeV and $1.7$ GeV \cite{Hilger:2015hka}. An RL-BSE study with beyond the leading order truncation showed the presence of a hybrid isovector resonance with mass $1.2 - 1.4$ GeV \cite{Qin:2011xq}. Another study modeling the hybrids by a two-particle interaction between the constituents was carried out in Ref. \cite{Xu:2018cor}. The Faddeev equation generated for the hybrids was solved in the RL-BSE scheme which yielded a mass of $\sim1.75(8)$ GeV. Curiously, this study interpreted (based on the nature of interactions used) the hybrid state as ``highly correlated $[gq][\bar{q}]\leftrightarrow q[g\bar{q}]$ bound state", where $[\ldots]$ represented the two-body channels \cite{Xu:2018cor}.\par

On the lattice front, the first attempts to study the hybrid mesons were first done by the UKQCD and MILC collaborations \cite{Lacock:1996vy,Lacock:1996ny,MILC:1997usn}. The MILC collaboration extracted the mass of the light $1^{-+}$ hybrid as $\sim 1.98$ GeV using the quenched lattice approximation \cite{MILC:1997usn}. In these calculations, the mixing of the hybrids with the tetraquarks were taken into account, but their effects were suppressed due to the quenching effects \cite{MILC:1997usn}. The UKQCD collaboration performed similar studies but without the mixing with the tetraquark configuration to arrive at identical predictions \cite{Lacock:1996vy,Lacock:1996ny}. A later study by the UKQCD collaboration by excluding the strange quarks resulted in an enhancement of the mass of the hybrids to $\sim 2.2$ GeV. Its width was estimated at $\sim 500$ MeV \cite{McNeile:2006bz}. More detailed studies were performed by the Hadron Spectrum (HadSpec) collaboration with the pion mass ranging from $\sim400$ MeV to $\sim700$ MeV and using dynamical anisotropic lattices as well as the distillation techniques \cite{Dudek:2010wm,Dudek:2011bn,Dudek:2011tt,Dudek:2013yja}. These studies gave a mass of $1.7 - 2$ GeV for the hybrid isovector \cite{Dudek:2010wm,Dudek:2011tt}, while simultaneously predicting the masses of the isoscalars to be larger than $2$ GeV \cite{Dudek:2011tt,Dudek:2013yja}. The most recent lattice calculations from the HadSpec collaboration involved a coupled channel calculation of the mass and decays of the lightest hybrid isovector. This study, the most accurate and comprehensive study so far, gave the $\pi_1(1600)$ a mass of $1564$ MeV and width in the range of $139-590$ MeV \cite{Woss:2020ayi}. Furthermore, the possible ranges for the partial widths of the decays of the $\pi_1(1600)$ into the $b_1\pi$, $\rho\pi$, $K^*K$, $f_1^{(\prime)}\pi$, $\eta^{(\prime)}\pi$, and $\rho\omega$ channels were also predicted \cite{Woss:2020ayi}.\par

In short, the theoretical studies of the hybrids using a variety of tools and techniques have resulted in a wide range of values for their masses and widths. However, the common theme underlying all these studies was the presence of one light hybrid meson with mass below $2$ GeV. The early works on the existence of exotic states in the light meson sector were vindicated when the first isovector resonance was observed in 1997 by the E852 experiment in the $\pi^- p\to \eta\pi^-p$ reaction \cite{E852:1997gvf}. A fit to the experimental data indicated the presence of a hybrid state with a mass of $1370\pm16 ^{+50}_{-30}$ MeV and a width of $385\pm 40 ^{+65}_{-105}$ MeV. However, an analysis of the $\pi^- p\to \pi^+\pi^-\pi^-p$ reaction by the same collaboration resulted in the discovery of an additional state \cite{E852:1998mbq}. This state, currently known as the $\pi_1(1600)$, had the exotic quantum numbers of $I^G(J^{PC})=1^-(1^{-+})$, a mass of $1593\pm 8 ^{+29}_{-47}$ MeV and a width of $168\pm 20^{+150}_{-12}$ MeV \cite{E852:1998mbq}. This analysis was subsequently expanded to a larger data set and the results were confirmed in Ref. \cite{Dzierba:2005jg}. The former, listed in the PDG as the $\pi_1(1400)$, was further confirmed by various experiments \cite{IHEP-Brussels-LosAlamos-AnnecyLAPP:1988iqi,CrystalBarrel:1998cfz,CrystalBarrel:1999reg,OBELIX:2004oio,E862:2006cfp,CrystalBarrel:2019zqh}. The latter was also confirmed by various experiments \cite{E852:2001ikk,E852:2004gpn,E852:2004rfa,COMPASS:2009xrl,COMPASS:2018uzl}. This lead to a puzzling situation were there were more hybrids than predicted. An analysis of the COMPASS data on the $\pi^- p\to \pi^+\pi^-\pi^-p$ scattering \cite{COMPASS:2018uzl} by the JPAC collaboration showed the presence of only one exotic resonance with a mass of $1564\pm24\pm86$ MeV and width of $492\pm54\pm102$ MeV \cite{JPAC:2018zyd}. It is currently believed that the additional (lighter) resonance observed by the experiments could be due to the interference of the background processes or final state interactions \cite{JPAC:2018zyd,Bass:2001zs}. \par

The current study is two-fold: on one hand we study the production of the hybrids via radiative decays of the $J/\psi$ \cite{Huang:2010dc,Page:1996ss,Rodas:2021tyb}, and on the other, we study the radiative decays of the hybrids themselves. The paper is organized as follows. In Sec. \ref{sec:mod} we provide the details of the model we have used to study the hybrid mesons. We also provide the methodology used to arrive at the results. In Sec. \ref{sec:RnD}, we present the results and discuss their implications. Finally, we summarize the paper in Sec. \ref{sec:SnC}.

\section{Model\label{sec:mod}}
\begin{figure}[t]
\centering
\includegraphics{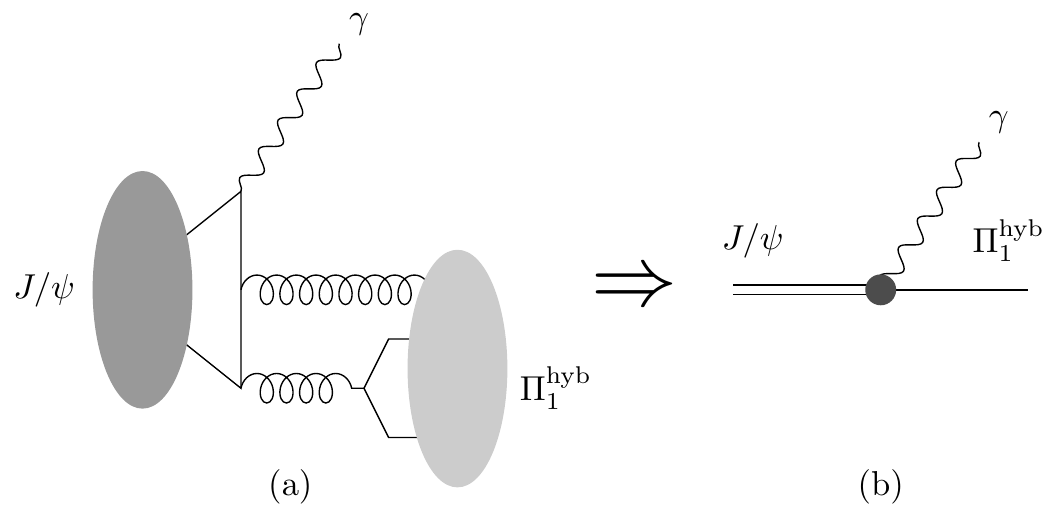}
\caption{(a) Schematic representation of the radiative production of the $1^{-+}$ hybrid. The shaded ellipses represent the $J/\psi$ and the hybrid. The internal lines represent the (anti)quarks (straight) and gluons (curly). (b) the Feynman diagram modeling the radiative production. The black filled circle represents the model vertex.}
\label{fig:feyndiarprd}
\end{figure}

\subsection{The hybrid nonet(s)}
We begin by giving a brief account of the constitution of the hybrids and their transformations under the chiral and discrete symmetries. The dynamics of light hadrons is dictated by the $SU(3)_L\times SU(3)_R \times U(1)_A$ symmetry. Hadrons of a given spin ($J$) and parity ($P$) can be constructed from a particular quark and/or gluon current. Based on this, to construct a mesonic state with $J^{PC}=1^{-+}$, a mere $\bar{q}q$ combination would be insufficient. We would thus have to include either a gluonic degree of freedom, or consider other exotic combinations like a tetraquark state or a meson molecule. The nonrelativistic quark models as well as the lattice studies have predicted one light hybrid nonet with mass below $2$ GeV. Further, lattice studies have shown that the isovector of the nonet has a mass and width similar to that of the $\pi_1(1600)$. Thus, it would be intuitive to consider the $\pi_1(1600)$ as a hybrid meson as opposed to a tetraquark or a molecular state.\par

The simplest current that can give rise to a $J^{PC}=1^{-+}$ state ($\Pi_{1,\mu}^\text{hyb}$) would be the one where the (anti)quarks and gluons interact via a vector current. This state is expected to have a chiral partner state ($B_{1,\mu}^\text{hyb}$) with $J^{PC}=1^{+-}$. The currents that make up the two states are given by,
\begin{align}
    \Pi^\text{hyb}_{1,\mu} &\to \frac{1}{\sqrt{2}}\bar{q} G_{\mu\nu} \gamma^\nu q\\
    B^\text{hyb}_{1,\mu} &\to \frac{1}{\sqrt{2}}\bar{q} G_{\mu\nu} \gamma^\nu \gamma_5 q
\end{align}
where, $G^{\mu\nu}_a=\partial^\mu A^\nu_a - \partial^\nu A^\mu_a - f_{abc}[A^\mu_b,A^\nu_c]$ is the gluonic field strength tensor and in the expressions above, the color indices have been suppressed. These two nonets are the so called ``homo-chiral'' nonets as the left ($L_\mu^\text{hyb}$) and right ($R_\mu^\text{hyb}$) chiral currents constructed from the hybrids transform under the chiral symmetry as \cite{Eshraim:2020ucw,Giacosa:2016hrm},
\begin{align}
    L_\mu^\text{hyb} \to U_L^\dagger L_\mu^\text{hyb} U_L &\,, ~~~ R_\mu^\text{hyb} \to U_R^\dagger R_\mu^\text{hyb} U_R
\end{align}
where,
\begin{align}
    L_\mu^\text{hyb} = \Pi^\text{hyb}_{1,\mu} + B^\text{hyb}_{1,\mu} &\,, ~~~ R_\mu^\text{hyb} = \Pi^\text{hyb}_{1,\mu} - B^\text{hyb}_{1,\mu}\text{ .}
\end{align}
The nonets arising out of these currents are given by,
\begin{align}
    \Pi^\text{hyb}_{1,\mu}=\frac{1}{\sqrt{2}}\begin{pmatrix}
    \frac{\eta_{1,N} +\pi_1^{0}}{\sqrt{2}} & \pi_1^{+} & K_1^+\\
    \pi_1^- & \frac{\eta_{1,N} - \pi_1^{0}}{\sqrt{2}} & K_1^0\\
    K_1^- & \bar{K}_1^0 & \eta_{1,S}
    \end{pmatrix}^\text{hyb}_{\mu},~~ &
    B^\text{hyb}_{1,\mu}=\frac{1}{\sqrt{2}}\begin{pmatrix}
    \frac{h_{1,N} +b_1^{0}}{\sqrt{2}} & b_1^{+} & K_{1,B}^+\\
    b_1^- & \frac{h_{1,N} - b_1^{0}}{\sqrt{2}} & K_{1,B}^0\\
    K_{1,B}^- & \bar{K}_{1,B}^0 & h_{1,S}
    \end{pmatrix}^\text{hyb}_{\mu}\text{ .}
\end{align}
The masses and some decays of these two nonets were studied earlier within the framework of the extended linear sigma model (eLSM) \cite{Eshraim:2020ucw}. This study identified the $\pi_1(1600)$ state as the isovector belonging to the $\Pi_{1,\mu}^\text{hyb}$ nonet. In Ref. \cite{Shastry:2022mhk}, the newly observed $\eta_1'(1855)$ was identified as the heavy isoscalar of the same nonet, and based on the available experimental data and lattice predictions, the two-body strong decays of $\pi_1(1600)$ and the $\eta_1'(1855)$ were studied. In addition, we predicted the mass and decay widths of the kaon and the light isoscalar hybrid meson. According to that analysis, the light isoscalar is expected to have a mass similar to that of the $\pi_1(1600)$ but a much smaller width, where as the kaon is expected to have a mass in the range of $1.7 - 1.8$ GeV and be as wide as the $\pi_1(1600)$ \cite{Shastry:2022mhk}. Because of the homo-chiral nature, the mixing between the isoscalars in these nonets are expected to be small \cite{Giacosa:2017pos}. However, absence of data on the partial widths of the $\eta_1'(1855)$ makes it difficult to pin-point the value of the mixing angle, thus we shall test also the effect of a nonzero value. Unfortunately, no members from the $B_{1,\mu}^\text{hyb}$ have been observed so far, and hence only approximate estimates can be made (see Ref. \cite{Eshraim:2020ucw} for details).
Here, we shall neglect this nonet of hybrid mesons in the following.\par

We now describe the models used to study the radiative production and decays of the hybrids.\par

\subsection{Radiative production of the light hybrid isoscalars}
The radiative production process we aim to study in the present work is schematically described in the Fig. \ref{fig:feyndiarprd}. We can write the Lagrangian for the radiative production of the hybrids as,
\begin{align}
    \mathcal{L}_{RP} &= g_{\gamma\eta_1} \psi_\mu F^{\mu\nu} \text{Tr}[\Pi^\text{hyb}_{1,\nu}],\label{eq:lagRP3}
\end{align}
where $\psi_\mu$ represents the vector charmonium, and $F^{\mu\nu}$ is the electromagnetic field strength tensor. The conservation of isospin restricts the production process to only the isoscalars. Thus, the Lagrangian can be explicitly written as,
\begin{align}
    \mathcal{L}_{RP} &= g_{\gamma\eta_1}\mathcal{C}_{\eta_L}\psi_\mu\eta_{1,\nu}^L F^{\mu\nu} + g_{\gamma\eta_1}\mathcal{C}_{\eta_H}\psi_\mu\eta_{1,\nu}^H F^{\mu\nu},\label{eq:lagRPind}
\end{align}
where $\mathcal{C}_i$ are coefficients depending on the mixing angle and the symmetry factors, and the light (heavy) isoscalar field is represented by $\eta_{1,\mu}^{L(H)}$.\par
\begin{table}[tb]
    \centering
    \begin{tabular}{|c|c|c|}
        \hline
        $\theta_h$ & \multicolumn{2}{c|}{$g_{\gamma\eta_1}$}\\\cline{2-3}
        & Set-1 & Set-2 \\\hline
        $0^\circ$ & $(1.07\pm 0.15)\times 10^{-2}$ & $(0.93\pm0.13)\times 10^{-2}$ \\\hline
        $15^\circ$ & $(1.78\pm 0.25)\times 10^{-2}$ & $(1.40\pm0.20)\times 10^{-2}$ \\\hline
    \end{tabular}
    \caption{The values of the coupling constant $g_{\gamma\eta_1}$ for two different $\theta_h$ (see discussion in here and in Ref. \cite{Shastry:2022mhk} for more details).}
    \label{tab:parval}
\end{table}

We now take a moment to analyse the nature of the interactions present in the Lagrangian. All the degrees of freedom involve in the Lagrangian have $J^P=1^-$. Note, the radiative production of the light hybrids can proceed through two angular momentum channels ($\ell=1,3$), with the $\ell=1$ channel receiving contributions from all possible spin configurations of the combined photon-hybrid system\footnote{Both these partial waves are present in the Lagrangian given in Eq. (\ref{eq:lagRP3},\ref{eq:lagRPind}). A point of concern would be the relative contributions of these two partial waves, particularly because the 3-momentum carried by the hybrid isoscalars could be as large as $1$ GeV. It is known from the previous analyses that the higher partial waves contribute significantly to the decay processes where the decay products carry large 3-momenta and to explain such large contributions, one needs higher order interactions in the Lagrangian \cite{Shastry:2021asu}. However, since not enough data is available to fix the coupling constants arising in such an elaborate Lagrangian, we restrict the analysis to lowest order terms.}. 
\begin{table}[tb]
    \centering
    \begin{tabular}{|c|c|c|}
        \hline
        State & \multicolumn{2}{c|}{Mass (MeV)}\\\hline
        & $\theta_h=0^\circ$ & $\theta_h=15^\circ$ \\\hline
        $\pi_1(1600)$ & \multicolumn{2}{c|}{$1661$} \\\hline
        $K_1^{hyb}(1750)$ & $1761$ & $1754$ \\\hline
        $\eta_1^{hyb}(1660)$ & $1661$ & $1646$\\\hline
        $\eta_1^\prime(1855)$ & \multicolumn{2}{c|}{$1855$}\\\hline
    \end{tabular}
    \caption{The masses of the hybrids for two different $\theta_h$ (see discussion in here and in Ref. \cite{Shastry:2022mhk} for more details).}
    \label{tab:massval}
\end{table}

The decay width for the process represented by the Feynman diagram in Fig. \ref{fig:feyndiarprd}(b) is given by,
\begin{align}
    \Gamma_{J/\psi\to\gamma\eta_1}(s) &= |g_{\gamma\eta_1}\mathcal{C}_{\eta_i}|^2\frac{k}{24~\pi~ m_J^2}\frac{(m_J^2+s)(m_J^2-s)^2}{2~s~m_J^2},\label{eq:dwRP}
\end{align}
where $s$ represents the square of the mass of the isoscalar, $m_J$ is the mass of $J/\psi$, and $k$ is the 3-momentum carried by the hybrid. We preserve the masses of the isoscalar hybrids as variables, as we shall be performing a spectral integration of these states.
\begin{figure}[t]
\centering
\includegraphics{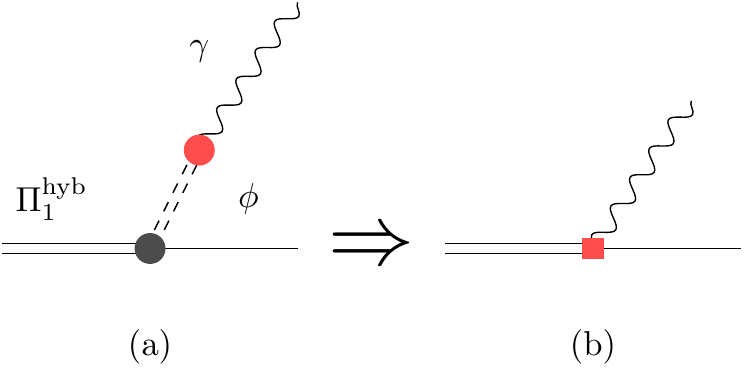}
\caption{(a) VMD picture of the radiative decay of the $1^{-+}$ hybrid into conventional states (represented as $\phi$); (b) the Feynman diagram modeling the radiative decay. The black and red blobs in (a) represent the FCP vertex and VMD vertex respectively, and the dashed double line represents a vector meson propagator. The red square in (b) represents the effective vertex.}
\label{fig:my_label}
\end{figure}

Namely, the radiative production of the hybrids involves the radiative decay of $J/\psi$ to the hybrid and the subsequent decay of the hybrid to two conventional mesons. The decay is thus a (two-step) sequential process and has to be evaluated accordingly \cite{PDGRev:2022abc}. Consider the following sequential decay process,
\begin{equation}
    A \to \mathcal{R}^*C_2 \to B C_1 C_2
\end{equation}
where, $A$ is the parent state (in our case, the $J/\psi$), $\mathcal{R}^*$ is the resonant intermediate state (hybrid meson), and $B$, $C_{1,2}$ are the final states ($\phi_{1,2}$ an $\gamma$ respectively). To arrive at the width of sequential decays, we calculate the partial width of the first step in the process ($A\to \mathcal{R}^*C_2$) and perform a spectral integration over the hybrid resonance ($\mathcal{R}^*$) to arrive at the partial width of the full process. The partial width of the full process is given by (e.g. Ref. \cite{Giacosa:2007bn}),
\begin{align}
    \Gamma_{A\to BC_1C_2} &= \int_{s_\text{th}}^{(\Delta M_{AC_2})^2} ds~ \Gamma_{A\to\mathcal{R}^*C_2}(s) d_s^i(s)\label{eq:specint}
\end{align}
where, $\sqrt{s}$ represents the mass of the unstable intermediate resonance, $\Gamma_{A\to\mathcal{R}^*C_2}(s)$ gives the partial width of $A\to\mathcal{R}^*C_2$ decay, and $d_s^i(s)$ is the spectral distribution function of the resonance as measured in the $i^\text{th}$ decay channel. Caution should be exercised in choosing the limits of spectral integration as the ``mass'' of the resonance cannot exceed the difference between the masses of the parent state ($A$) and the decay partner ($C_2$) ($\Delta M_{AC_2}$). This is inherently taken care of by the 3-momentum of the resonance present in the phase-space part of the decay width. Further, in case of a Breit-Wigner-like distribution, the lower limit of the integration must represent the decay threshold ($s_\text{th}$).\par

In the present work, we use the recently proposed Sill-distribution \cite{Giacosa:2021mbz} to model the decay of the hybrid resonance to two conventional states. The Sill-distribution has the advantages of having a built-in lower threshold as well as normalization. Yet, it should be stressed that the results change only slightly if a different distribution (such as relativistic Breit-Wigner one) is employed, provided the threshold is not near by (as in the case of the $\pi_1(1600)$) \cite{Giacosa:2021mbz}. However, the difference would be stark, in case of the remaining members of the nonet as the thresholds of their dominant decay channels lie very close to the masses \cite{Giacosa:2021mbz,Shastry:2022mhk}. To take care of these possible complications we use the Sill distribution in the present work.\par

The single channel Sill-distribution is given by \cite{Giacosa:2021mbz},
\begin{align}
    d_s(s) &= \frac{1}{\pi}\frac{\sqrt{s-s_\text{th}}~\tilde{\Gamma
}}{(s-M^{2})^{2}+(\sqrt{s-s_\text{th}}~\tilde{\Gamma})^{2}}\theta(s-s_\text{th})\label{eq:specsing}
\end{align}
where, $M$ is the mass of the resonance, and if $\Gamma$ is the total width of the resonance, then,
\begin{align}
    \tilde{\Gamma}=\Gamma\frac{M}{\sqrt{M^{2}-s_\text{th}}}\text{ .}\label{eq:gtilde}
\end{align}
Further, the Sill-distribution can be extrapolated to study the line-shapes of the resonances with multiple decay channels. The Sill-distribution of such a system as measured in the $k^\text{th}$ channel is given by \cite{Giacosa:2021mbz},
\begin{align}
    d_s^k(s) &= \frac{1}{\pi}\frac{\sqrt{s-s_\text{th,k}}~\tilde{\Gamma
}_k}{(s-M^{2}-\sum_{i=1}^{Q}\sqrt{s_\text{th,i}-s}~\tilde{\Gamma}_i)^{2}+\sum_{i=Q+1}^{N}(\sqrt{s-s_\text{th,i}}~\tilde{\Gamma}_i)^{2}}\theta(s-s_\text{th,k})\label{eq:specmult}
\end{align}
where, $s_\text{th,k}$ is the $k^\text{th}$ threshold, $\tilde{\Gamma}_k$ is the quantity given in Eq. \ref{eq:gtilde} for the $k^\text{th}$ decay channel, and the integer $Q$ is such that, for all $i<Q$, $s_{th,i}<s_{th,k}$.
\begin{table*}[t]
\centering
    \begin{tabular}{|c|c|c|c|c|}
    \hline
    Production & \multicolumn{4}{c|}{Branching ratio ($10^{-4}$)}\\\cline{2-5}
    Channel & \multicolumn{2}{c|}{Set-1}& \multicolumn{2}{c|}{Set-2} \\\cline{2-5}
    ($\phi_1\phi_2$) & $\theta_h=0^\circ$ & $\theta_h=15^\circ$ & $\theta_h=0^\circ$ & $\theta_h=15^\circ$ \\\hline
    \multicolumn{5}{|c|}{$\eta_1^{hyb}(1660)$}\\\hline
    $a_1\pi$  & $4.8\pm1.4$ & $16.2\pm4.6$ & $3.8\pm1.1$ & $10.2\pm2.9$ \\\hline
    $K^*K$ & $(1.73 \pm0.49)\times 10^{-2}$ & $(2.35\pm0.67)\times 10^{-2}$ & $(1.29\pm0.37)\times 10^{-2}$ & $(1.38\pm0.39)\times 10^{-2}$ \\\hline
    $\eta^\prime\eta$ & $(2.28\pm0.65)\times 10^{-2}$ & $(13.5\pm 3.8)\times 10^{-2}$ & $(1.71\pm0.49)\times 10^{-2}$ & $(8.3\pm2.4)\times 10^{-2}$ \\\hline
    $\rho \rho$ & $(4.4\pm 1.3)\times 10^{-3}$ & $(13.1\pm3.8)\times 10^{-3}$ & $(3.36\pm0.96)\times 10^{-3}$ & $(8.3\pm2.4)\times 10^{-3}$ \\\hline
    \multicolumn{5}{|c|}{$\eta_1^\prime(1855)$}\\\hline
    $K_1(1270)K$ & $2.45\pm0.70$ & $2.82\pm0.80$ & $1.49\pm0.42$ & $1.29\pm0.37$ \\\hline
    $K^*K$ & $(1.86\pm0.53)\times 10^{-2}$ & $(1.90\pm0.54)\times 10^{-2}$ & $(1.83\pm0.52)\times 10^{-2}$ & $(1.86\pm0.53)\times 10^{-2}$ \\\hline
    $K^*K^*$ & $(7.2\pm2.1)\times 10^{-4}$ & $(8.0\pm2.3)\times 10^{-4}$ & $(7.6\pm2.2)\times 10^{-4}$ & $(8.6\pm2.4)\times 10^{-4}$ \\\hline
    $f_1(1285)\eta$ & $(27.6\pm7.9)\times 10^{-3}$ & $(1.75\pm0.50)\times 10^{-3}$ & $(27.9\pm8.0)\times 10^{-3}$ & $(1.82\pm0.52)\times 10^{-3}$ \\\hline
    $\eta\eta^\prime$ & \multicolumn{4}{c|}{$(2.70\pm0.76)\times 10^{-2}$\cite{BESIII:2022iwi}}\\\hline
    \end{tabular}
    \caption{The branching ratios of the $J/\psi\to\gamma\eta_1^{hyb}(1660)\to\gamma\phi_1\phi_2$ and $J/\psi\to\gamma\eta_1^\prime(1855)\to\gamma\phi_1\phi_2$ decays in the allowed tree-level decay channels of the respective hybrids \cite{Shastry:2022mhk}. The last row represents the value used to fix the parameters. The error estimates do not include uncertainties in the masses of the parent state.}\label{tab:pardecwidLH}
\end{table*}

\subsection{VMD and radiative decays of the light hybrids}
To study the radiative decays of the hybrids, we use the vector meson dominance model (VMD) in conjunction with the Lagrangian reported in Ref. \cite{Shastry:2022mhk}. The relevant part of the FCP Lagrangian describing the decay of the hybrids into at least one vector state is given by,
\begin{align}
    \mathcal{L}_{VX} &= g_{\rho\pi} \text{Tr}\left[\tilde{\Pi}_{1,\mu\nu}[P,V^{\mu\nu}]\right]+ g_{\rho\omega} \text{Tr}\Big[\Pi_{1,\mu}\{V^{\mu\nu},V_\nu\}\Big]
\end{align}
where $g_{\rho\pi}$ and $g_{\rho\omega}$ are the respective coupling constants. To obtain a description of the radiative decay process, we perform the following transformation \cite{Piotrowska:2017rgt} (cf., Fig. \ref{fig:my_label}),
\begin{align}
    V_{\mu\nu} &\to V_{\mu\nu} + \frac{e_0}{g_\rho}\mathcal{Q}F_{\mu\nu}
\end{align}
where, $\mathcal{Q}=\text{diag}\{2/3,-1/3,-1/3\}$, $e_0$ is the charge and $g_\rho$ is the coupling constant. The ratio $\frac{e_0}{g_\rho}$ originally described the coupling of the $\rho^0$ to an electron-positron pair \cite{OConnell:1995nse}. In the VMD picture, this ratio describes the coupling of the $\rho^0/\omega/\phi$ mesons to a photon. The Lagrangian describing the radiative decay of hybrids can thus be written as,
\begin{align}
    \mathcal{L}_{\gamma X} &= g_{\rho\pi}\frac{e_0}{g_\rho} \text{Tr} \left[\tilde{\Pi}_{1,\mu\nu}[P,\mathcal{Q}] \right] F^{\mu\nu} + g_{\rho\omega}\frac{e_0}{g_\rho} \text{Tr} \Big[ \Pi_{1,\mu}\{\mathcal{Q},V_\nu\}\Big]F^{\mu\nu}.\label{eq:LagRaddec}
\end{align}
Interestingly, the neutral hybrids are forbidden from decaying radiatively to pseudoscalar states (at the tree-level) in the above Lagrangian. This feature arises from the form of the interaction term representing such a decay process. In the present calculation, the $\rho^0\to e^+ e^-$ decay width is used to extract the value of $\frac{e_0}{g_\rho}$. Since the partial width for the $\rho^0\to e^+ e^-$ decay is $7.04 \pm 0.06\text{ keV}$ \cite{Workman:2022ynf}, we get,
\begin{align}
    \frac{e_0}{g_\rho} &= 0.0185\pm 0.00079.
\end{align}
We note here that the relatively small uncertainty in the value above has not been taken into account while estimating the uncertainties in the parameters of the current study or in those of the partial widths. The coupling constants $g_{\rho\omega}$ and $g_{\rho\pi}$ were estimated in Ref. \cite{Shastry:2022mhk}, and have the values $g_{\rho\pi}=0.35\pm0.05\text{ GeV}$ and $g_{\rho\omega}=-(0.37\pm0.07)$.\par

\begin{table*}[tbp]
\centering
    \begin{tabular}{|c|c|c|c|c|c|}
    \hline
    Decay & \multicolumn{2}{c|}{Width (keV)} & Decay & \multicolumn{2}{c|}{Width (keV)} \\ \cline{2-3}\cline{5-6}
    Channel & $\theta=15^\circ$ & $\theta=0^\circ$ & & $\theta=15^\circ$ & $\theta=0^\circ$\\\hline
    \multicolumn{3}{|c|}{$\pi_1(1600)$} & \multicolumn{3}{c|}{$\eta_1^{hyb}(1660)$}\\\hline
    $\pi \gamma$ & \multicolumn{2}{c|}{$4.7\pm1.4$}  &
    $\rho \gamma $ & $0.147\pm0.061$ & $0.164\pm0.068$ \\\hline
    $\rho \gamma $ & \multicolumn{2}{c|}{$(5.5\pm2.1)\times 10^{-2}$} & $\omega \gamma$ & $(1.67\pm0.69)\times 10^{-2}$ & $(1.76\pm0.73)\times 10^{-2}$ \\\hline
    $\omega \gamma $ & \multicolumn{2}{c|}{$0.158\pm0.060$} & $\phi \gamma $ & $(1.23\pm0.51)\times 10^{-3}$ & $(1.64\pm0.68)\times 10^{-5}$\\\hline
    $\phi \gamma$ & \multicolumn{2}{c|}{$(1.48\pm0.56)\times 10^{-4}$} & \multicolumn{3}{c|}{$\eta_1^\prime(1855)$} \\\hline
    \multicolumn{3}{|c|}{$K_1^{ hyb}(1750)$} & $\rho \gamma$ & $(1.73\pm0.72)\times 10^{-2}$ & $0.$\\\hline
    $K\gamma$ & $4.4\pm1.3$ & $4.5\pm1.3$ & $\omega \gamma $ & $(6.9\pm 2.9)\times 10^{-4}$ & $(3.1\pm1.27)\times 10^{-4}$ \\\hline
    $K^*\gamma$  & $0.142\pm 0.059$ & $0.145\pm0.060$ & $\phi \gamma $ & $(4.18\pm1.73)\times 10^{-2}$ & $(4.42\pm1.83)\times 10^{-2}$  \\\hline
    \end{tabular}
    \caption{Radiative decay widths of the $1^{-+}$ hybrids for two values of isoscalar mixing angle. Note that the radiative decay widths of the $\pi_1(1600)$ do not depend on the isoscalar angle, where as, those of the $K_1^{hyb}(1750)$ depend on the isoscalar mixing angle through the strangeness contribution to its mass (see text for details). The error estimates do not include uncertainties in the masses of the parent state.}\label{tab:raddecisv}
\end{table*}

\section{Results and Discussions\label{sec:RnD}}
We now present the results of the present study beginning with the radiative production of the hybrids. The results of these studies are presented in Table \ref{tab:pardecwidLH} and Table \ref{tab:raddecisv}.\par

A comment on the choice of the parameters and the mixing angle is in order. The values of the partial widths of the various decay channels of the hybrids used in the present work were taken from Ref. \cite{Shastry:2022mhk}. In arriving at these values we had made use of the ratio of the branching ratio of the $\pi_1(1600)\to b_1(1235)\pi$ decay in the $D$-wave ($\ell=2$) to that in the $S$-wave ($\ell=0$), in addition to the partial widths predicted by the lattice study \cite{Woss:2020ayi}. This ratio is equal to the square of the ratio of the partial wave amplitudes ($D/S$-ratio). It was argued that a large value for this ratio implied large contribution from the higher order interactions, and that the ratio of the corresponding coupling constants could be fixed from the $D/S$-ratio \cite{Shastry:2021asu}. However, since the sign of the ratio of the partial wave amplitudes is not known, the fit was performed for both cases, leading to two sets of coupling constants, and hence, two sets of partial and/or total widths. Since this ambiguity is still unresolved, we use both the sets of partial widths in this study. Accordingly, ``Set-1'' corresponds to the value when the said $D/S$-ratio is positive and ``Set-2'' when it is negative. This issue does not arise for the radiative decays of the hybrids as the partial widths for the parent decay (vector-pseudoscalar and vector-vector) channels do not vary (see, Ref. \cite{Shastry:2022mhk} for more details).\par

The uncertainty in the available data also affects our understanding of the extent of mixing between the isoscalars. As mentioned in Sec. \ref{sec:mod}, the light hybrids are homo-chiral in nature, implying a small mixing angle in the isoscalar sector. However, as it is known from the studies of various conventional mesons, the value of the mixing angle for a pair of homo-chiral isoscalars need not be zero \cite{Shastry:2021asu}. Thus, we assume two possible values for the mixing angle: $\theta_h=0^\circ$ and $\theta_h=15^\circ$. The latter value is representative of a small but non-zero mixing angle, and was chosen as an illustrative value to show the effect of a small but nonzero mixing angle\cite{Shastry:2022mhk}.

The error estimates in the partial widths for the radiative production arise primarily due to the uncertainties in the experimental input \cite{BESIII:2022iwi}. Those for the radiative decays are due to the uncertainties in the coupling constants $g_{\rho\pi}$ and $g_{\rho\omega}$. We also note here that the errors in the partial widths $\Gamma_i$ used in the spectral integration can in principle contribute to the uncertainties in the BRs for the radiative production of the hybrids. But, these uncertainties are estimated to contribute $\sim 2\%$ to the existing $\sim 30\%$ error coming from the data, and hence have not been included here (see, Appendix \ref{sec:appa} for more details).\par

Finally, we comment on the masses of the hybrid kaon and the light hybrid isoscalar which are as yet unobserved, and hence their masses have been estimated using flavor symmetry and axial anomaly constraints \cite{Shastry:2022mhk}. 
The strangeness contribution to the masses of these states was estimated from the mass of the heavy isoscalar and is a function of the mixing angle. We mention here the hybrid kaon masses depend indirectly on the isoscalar mixing angle. This is due to the fact that we have two parameters in our model (strangeness contribution and mixing angle) but only one data point (mass of the heavy hybrid isoscalar) \cite{Shastry:2022mhk}. The masses of the hybrid states are listed in Table \ref{tab:massval}.

\subsection{Radiative production}
The Lagrangian given in Eq. \ref{eq:lagRP3} has one independent parameter - the coupling constant $g_{\gamma\eta_1}$. This parameter can be estimated from the $J/\psi\to\gamma\eta_1'(1855)\to\gamma\eta\eta^\prime$ branching ratio reported by the BESIII collaboration \cite{BESIII:2022riz}. The resultant values are listed in Table \ref{tab:parval}. While extracting the value of this coupling constant, we have considered two possibilities for the mixing angle - $\theta_h=0^\circ$ and $\theta_h=15^\circ$. We have also used both the sets of parameters mentioned in Ref. \cite{Shastry:2022mhk}. We note here that the values of the coupling constant do not vary appreciably with the change in the strong decay coupling constants, as the difference in the partial widths of the $\eta\eta^\prime$ channel is not appreciably large \cite{Shastry:2022mhk}. However, there is a significant difference in the values of $g_{\gamma\eta_1}$ when the mixing angle is changed.\par

The branching ratios (BRs) for the various possible radiative decays of the $J/\psi$ via an intermediate hybrid isoscalar are presented in Table \ref{tab:pardecwidLH}. The BRs follow the same pattern as that of the partial widths of the isoscalars. In the case of the $\eta_1^{hyb}(1660)$, the BR for the $\gamma a_1(1260)\pi$ channel is the largest. The $\gamma K^*K$ and the $\gamma\eta\eta^\prime$ channels have nearly the same BRs when $\theta_h=0$. However, this pattern is broken when a small mixing is introduced in the form of $\theta_h=15^\circ$. In this case, the BR for the $\gamma K^*K$ channel decreases slightly, where as that for the $\gamma\eta^\prime\eta$ channel increases by nearly a factor of $6$. This behavior is valid for the second set of parameters as well.\par

The general features discussed above are retained by the $J/\psi \to \gamma\eta_1^\prime(1855)\to\gamma\phi_1\phi_2$ decays as well. The $\gamma K_1(1270)K$ channel has the largest BR, while for a given mixing angle the $\gamma K^*K$ and the $\gamma\eta f_1(1285)$ channels have similar BRs as that of the $\gamma\eta\eta^\prime$ channel where the $\eta_1^\prime(1855)$ has been observed. The $\gamma\eta f_1(1285)$ channel is a rather interesting channel in that the $f_1(1285)$ state is a relatively stable state that can be observed in the experiments. Further, according to our calculations, this channel is very much sensitive to the hybrid isoscalar mixing angle. The BR varies by a factor of $\sim 15$ when the mixing angle is increased to $\theta_h=15^\circ$ from $\theta_h=0^\circ$. This variation can be attributed to the fact that the $\eta_1^\prime(1855)\to f_1(1285)\eta$ decay involves three isoscalars - the $\eta_1^\prime(1855)$, the $ f_1(1285)$, and the $\eta$. If the angle of the $f_1-f_1^\prime$ mixing is given by $\theta_a$, and that of the $\eta-\eta^\prime$ by $\theta_p$, then the decay width contains the combination,
\begin{align}
& \left(\frac{1}{\sqrt{2}}\sin (\theta_a) \cos (\theta_h) \sin
   (\theta_p)-\frac{1}{2} \cos (\theta_a)
   \sin (\theta_h) \cos (\theta_p)\right)^2.
\end{align}
When $\theta_a=24^\circ$ \cite{LHCb:2013ged}, and $\theta_p=-44.5^\circ$ \cite{Amelino-Camelia:2010cem}, the value of the mixing contribution varies by a factor of $2$ when the hybrid mixing angle is increased from $\theta_h=0^\circ$ to $\theta_h=15^\circ$ and its influence can be seen in the different partial widths for the $\eta_1^\prime(1855)\to f_1\eta$ decay \cite{Shastry:2022mhk}. It is this effect that carries over to the radiative decays.
This channel could thus provide valuable information regarding the mixing between the isoscalars, which could then be used to narrow down on the mass and width of the unknown light isoscalar hybrid - the $\eta_1^{hyb}(1660)$. Finally, summing all the channels, we find that the BR for the $J/\psi\to\eta_1(1600)$ decay is of the order of $10^{-3}-10^{-4}$, where as that for the $J/\psi\to\eta_1^\prime(1855)$ decay is of the order of $10^{-4}$.\par

The $\pi_1(1600)$ cannot be readily observed in the radiative decays of the heavy quarkonia as such decays would break the isospin symmetry. However, one could in principle study processes involving the $\rho$-meson instead of photon as suggested in Ref. \cite{Huang:2010dc}. Such a production process would contain dominant contributions from gluon mediated production of the vector meson, in addition to the suppressed photon mediated production. Such a study will be attempted elsewhere.

\subsection{Radiative decays}
We now discuss the radiative decays of the hybrids. The VMD $+$ FCP Lagrangian given in Eq. \ref{eq:LagRaddec} shows that the hybrids can decay radiatively to the pseudoscalar mesons and the vector mesons. The underlying mechanisms for the two decays are significantly different as evident from the form of the Lagrangian. This difference is manifest in the allowed decays listed in Table \ref{tab:raddecisv}.
In particular, the isoscalar hybrids cannot decay radiatively to pseudoscalar states in the tree-level. This is also true for the neutral $\pi_1(1600)$ and the $K_1^{hyb}(1750)$. Further, the hybrids cannot decay radiatively to the isoscalar pseudoscalars ($\eta^{(\prime)}$). These constraints do not apply to the vector meson states. We also note here that the branching ratios for the radiative decays of the hybrids are typically $\lesssim 10^{-5}$. Moreover, the VMD transformation opens up new channels for the decay of the hybrids which are kinematically forbidden in the corresponding hadronic channels like, the $(\omega/\phi)\gamma$ channel for the decays of the non-kaonic hybrids.\par

The partial widths of the radiative decays of the light isoscalar $\eta_1^{hyb}(1660)$ are nearly independent of the mixing angle except for the $\phi\gamma$ channel, where as those of the heavy isoscalar $\eta_1^\prime(1855)$ do depend on the value of the mixing angle. Specifically, at the tree-level, the $\eta^\prime(1855)\to\rho\gamma$ decay is forbidden if the mixing angle is $\theta_h=0^\circ$. The partial width for the $\eta^\prime(1855)\to\omega\gamma$ decay changes by a factor of $\sim 2$ when the mixing angle is increased from $\theta_h=0^\circ$ to $\theta_h=15^\circ$. The partial width for the $\eta_1^\prime(1855)\to\phi\gamma$ decay is, however, independent of the hybrid mixing angle. The smallness of the hybrid mixing angle also implies that the $\rho\gamma$ channel is largest decay channel for the radiative decay of the light isoscalar $\eta_1^{hyb}(1660)$, while the $\phi\gamma$ channel is the smallest among those allowed at the tree-level. The converse is true for the heavier $\eta_1^\prime(1855)$ where the $\phi\gamma$ channel is the largest and the $\omega\gamma$ channel is the smallest. Intuitively, this could be attributed to the fact that the $\omega$ and $\phi$ states are nearly pure $\bar{n}n$ and $\bar{s}s$ states, and hence the small mixing angle of the hybrids cannot result in appreciably large $\bar{s}s$ content in the $\eta_1(1600)$ and vice versa.\par

Thus, the observation of any one of the hybrid isoscalars in a radiative channel can shed important light on the extent of mixing between them. 

\section{Summary and Conclusions\label{sec:SnC}}
In this paper, we have presented the results of our model based study of the radiative production and decays of the light ($1^{-+}$) hybrid mesons. We have studied the production of the hybrid mesons via the radiative decay of the $J/\psi$ using the $J/\psi \to \gamma \eta_1^\prime(1855)\to\gamma\eta\eta^\prime$ process observed by the BESIII collaboration as the reference \cite{BESIII:2022riz}. We find that the $\gamma\eta f_1(1285)$ channel is very sensitive to the angle of mixing between the hybrid isoscalars in addition to the $\gamma\eta\eta^\prime$ final state. Even though the effects of the mixing are present in the $\gamma\pi a_1(1260)$ channel, the unstable nature of the $a_1(1260)$ makes this channel quite challenging for experimental studies. Furthermore, the branching ratios for the radiative decays of the $J/\psi$ to the hybrid isoscalars are expected to be of the order of $10^{-3}-10^{-4}$, which is nearly an order of magnitude smaller than the branching ratios for the $J/\psi\to\gamma\eta^{(\prime)}$ decays.\par

The radiative decays of the hybrid mesons are governed by the vector meson dominance. Based on our study of the strong decays of the hybrids, we find that the hybrids can decay into pseudoscalar mesons and vector mesons radiatively. However, the number of pseudoscalar channels available for the decays are limited by the conservation of the isospin and $G$-parity. In particularly, the neutral hybrids cannot decay into any pseudoscalar mesons radiatively. We find that the vector meson decay channels of the isoscalars are very much sensitive to the mixing between the hybrid isoscalars and can be used to understand the quark content of these states.

\section*{Acknowledgements}
The authors acknowledge financial support from the Polish National Science Centre (NCN) via the OPUS project 2019/33/B/ST2/00613.

\appendix
\section{Partial widths and error estimates}\label{sec:appa}

In this appendix, we briefly recall the fit performed in Ref. \cite{Shastry:2022mhk} and discuss the error estimates used in the present work.\par

In Ref. \cite{Shastry:2022mhk}, a model Lagrangian invariant under $SU(3)$ flavor symmetry, parity reversal, and charge conjugation was used to describe the two-body decays of the light hybrids. The parameters of the Lagrangian were fitted to the available data which included the mass and full width of $\pi_1(1600)$ \cite{Workman:2022ynf}, the $D/S$-ratio for the $\pi_1(1600)\to b_1\pi$ decay \cite{Workman:2022ynf}, the lattice estimates for the partial widths of the decays of the $\pi_1(1600)$ \cite{Woss:2020ayi}, and flavor symmetry constraints. The lattice estimates provided only possible ranges for the partial widths. For this reason, the midpoint of those ranges were used with $50\%$ uncertainties. The flavor constraints were assumed to have $30\%$ uncertainties. A $\chi^2$-fit was performed using these data to arrive at the values of the parameters listed in Table 2 of Ref. \cite{Shastry:2022mhk}. The uncertainties in the values of the parameters as well as in the values of the partial widths were estimated using the Hesse matrix formalism \cite{Shastry:2022mhk}.\par

In Table \ref{tab:parlist1}, we list the values of the partial widths used in arriving at the results in Table \ref{tab:pardecwidLH}.

\begin{table}[h]
    \centering
    {\renewcommand{\arraystretch}{1.5}
    \begin{tabular}{|c|c|c|c|c|}
    \hline
    Decay & \multicolumn{2}{c|}{Set-1}& \multicolumn{2}{c|}{Set-2} \\\cline{2-5}
    Channel & $\theta_h=15^\circ$ & $\theta_h=0^\circ$ & $\theta_h=15^\circ$ & $\theta_h=0^\circ$\\\hline
    \multicolumn{5}{|c|}{$\eta_1^{hyb}(1660)$}\\\hline
    $a_1\pi$ & $68\pm13 $ & $80\pm15$ & $71\pm14 $ & $82\pm16 $ \\\hline
    $K^*K$ & $0.10\pm0.026 $ & $0.29\pm0.075 $ & $0.10\pm0.026 $ & $0.29\pm0.075 $ \\\hline
    $\eta^\prime\eta$ & $0.61\pm0.13 $ & $0.41\pm0.09 $ & $0.62\pm0.14 $ & $0.41\pm0.09 $ \\\hline
    $\rho\rho$ & $0.060\pm0.021 $ & $0.081\pm0.028 $ & $0.062\pm0.021 $ & $0.082\pm0.028 $ \\\hline
    \multicolumn{5}{|c|}{$\eta_1^\prime(1855)$}\\\hline
    $K_1(1270)K$ & $392\pm142 $ & $253\pm92 $ & $172\pm76 $ & $151\pm67 $ \\\hline
    $K^*K$ & $1.92\pm0.50 $ & $1.45\pm0.37 $ & $1.93\pm0.50 $ & $1.46\pm0.38 $ \\\hline
    $\eta^\prime\eta$ & $2.97\pm0.66 $ & $2.28\pm0.51 $ & $3.02\pm0.67 $ & $2.31\pm0.51 $ \\\hline
    $K^*K^*$ & $0.11\pm0.040 $ & $0.075\pm0.027 $ & $0.12\pm0.044 $ & $0.077\pm0.028 $ \\\hline
    $f_1(1285)\eta$ & $0.18\pm0.046 $ & $2.15\pm0.56 $ & $0.19\pm0.034 $ & $2.21\pm0.57 $ \\\hline
    \end{tabular}}
    \caption{Values of the partial widths used in arriving at the results listed in Table \ref{tab:pardecwidLH}.}
    \label{tab:parlist1}
\end{table}

The errors in the values listed in Table \ref{tab:parlist1} contribute to the uncertainties in the BRs for the radiative production of the hybrid isoscalars in the respective channels via Eq. \ref{eq:specmult}. To arrive at an estimate of these contributions, we use the derivative approach given by,
\begin{align}
    \Delta \Gamma_\text{RP}^i &= \sum_i \int_{s_{th}}^\infty ds \frac{\partial}{\partial\alpha_i}\left(d_s^i(s)\Gamma_{J/\psi\to\gamma\eta_1}^i(s)\right)\Delta\alpha^i
\end{align}
where, $\alpha^i$ is the partial width of the $i^\text{th}$ decay channel of the hybrid and $\Delta\alpha^i$ is its uncertainty. Using the values listed in Table \ref{tab:parlist1}, we find that the total contribution is of the order of $1-2\%$. This value could be further diminished by using the Hesse matrix approach  (for an brief review of this method in the realm of mesons, see e.g. \cite{Piotrowska:2017rgt}), 
but, since it is already small, 
 we can safely ignore the uncertainties from the partial widths of the hybrid isoscalars.
 \par

\end{document}